# *In-situ* crystallographic mapping constrains sulfate deposition and timing in Jezero crater, Mars

## Crystal texture mapping of sulfates on Mars

**Authors**

Michael W. M. Jones[1,2,3]*, David T. Flannery[3,4], Joel A. Hurowitz[5], Mike T. Tice[6], Christoph E. Schrank[3,4], Abigail C. Allwood[7], Nicholas J. Tosca[8], David C. Catling[9], Scott J. VanBommel[10], Abigail L. Knight[10], Briana Ganly[11], Kirsten L. Siebach[12], Kathleen C. Benison[13], Adrian P. Broz[14], Maria-Paz Zorzano[15], Chris M. Heirwegh[7], Brendan J. Orenstein[3,4], Benton C. Clark[16], Kimberly P. Sinclair[9], Andrew O. Shumway[9], Lawrence A. Wade[7], Scott Davidoff[7], Peter Nemere[3,4], Austin P. Wright[17], Adrian E. Galvin[7], Nicholas Randazzo[18], Jesús Martinez-Frias[19], Lauren P. O'Neil[6]

**Affiliations**

[1]Central Analytical Research Facility, Queensland University of Technology, Brisbane, Australia, 4000.
[2]School of Chemistry and Physics, Queensland University of Technology, Brisbane, Australia, 4000.
[3]Planetary Surface Exploration Group, Queensland University of Technology, Brisbane, Australia, 4000
[4]School of Earth and Atmospheric Sciences, Queensland University of Technology, Brisbane, Australia, 4000.
[5]Department of Geosciences, Stony Brook University, Stony Brook, NY 11794, USA.
[6]Department of Geology and Geophysics, Texas A&M University, College Station, TX 77843, USA.
[7]Jet Propulsion Laboratory, California Institute of Technology, Pasadena, CA 91109, USA.
[8]Department of Earth Sciences, University of Cambridge, Cambridge, UK.
[9]Department of Earth and Space Sciences, University of Washington, Seattle WA 98195, USA.
[10]Department of Earth, Environmental, and Planetary Sciences, Washington University in St. Louis, St. Louis, MO, USA, 63130.
[11]Mineral Resources, Commonwealth Scientific and Industrial Research Organisation, Sydney, NSW, Australia.
[12]Department of Earth, Environmental and Planetary Sciences, Rice University, Houston, TX, USA 77005.
[13]Department of Geology and Geography, West Virginia University, Morgantown, WV, USA, 26506.
[14]Purdue University, Department of Earth, Atmospheric and Planetary Sciences, West Lafayette, IN, USA, 47907.
[15]Centro de Astrobiología (CAB), CSIC-INTA, 28850 Torrejón de Ardoz, Madrid, Spain.
[16]Space Science Institute, Boulder, CO 80301, USA.
[17]School of Computational Science and Engineering, Georgia Institute of Technology, Atlanta, GA 30332, USA.
[18]Department of Earth and Atmospheric Sciences, University of Alberta, Edmonton, Alberta, Canada.
[19]Institute of Geosciences, CSIC-UCM, Spain.

[Type here]

*Corresponding author. Email: mw.jones@qut.edu.au

**Abstract**

Late-stage Ca-sulfate-filled fractures are common on Mars. Notably, the Shenandoah formation in the western edge of Jezero crater preserves a variety of Ca-sulfate minerals in the fine-grained siliciclastic rocks explored by the *Perseverance* rover. However, the depositional environment and timing of the formation of these sulfates is unknown. To address this outstanding problem, we developed a new technique to map the crystal textures of these sulfates *in situ* at two stratigraphically similar locations in the Shenandoah formation, allowing us to constrain the burial depth and paleoenvironment at the time of their deposition. Our results suggest that some Ca-sulfate analyzed was formed at a burial depth greater than 80m, whereas Ca-sulfates present at another outcrop likely precipitated in a shallow-subsurface environment. These results indicate that samples collected for potential return to Earth at the two studied locations capture two different times and distinct chemical conditions in the depositional history of the Shenandoah formation providing multiple opportunities to evaluate surface and subsurface habitability.

**Teaser**

Crystal texture analysis of Ca-sulfates on Mars indicates deposition at varying times covering multiple paleoenvironments.

## MAIN TEXT

### Introduction

Since landing in Jezero crater on Mars in February, 2021, the Mars 2020 *Perseverance* rover has explored igneous rocks exposed in the crater floor (*1, 2*) and sedimentary rocks preserved in paleolake and deltaic environments (*3*). The overarching goals of the Mars 2020 mission are to constrain the habitability of these paleoenvironments, to search for biosignatures (*4*) and to collect a compelling cache of samples for potential Mars Sample Return (*5*).

Fine-grained clastic rocks were analyzed by *Perseverance* as the rover traversed the main sedimentary fan exposed in the western edge of Jezero crater, in a succession informally known as the "Shenandoah formation" (*6*). Two study areas in this region, at "Hogwallow Flats" ("Wildcat Ridge" outcrop) and at "Yori Pass" ("Hidden Harbor" outcrop), are notable for abundant sulfate minerals present in clasts, cements, vugs, and in veins (*7*). The Sample Caching System (*8*) collected three rock cores for potential return to Earth: "Hazeltop" and "Bearwallow" at Wildcat Ridge, and "Kukaklek" at Hidden Harbor (**Fig. S1**). These sulfates offer remarkable opportunities to investigate past environmental conditions and potentially also provide evidence of past life on Mars (*9*). However, the timing and context of the formation of these minerals is crucial to interpreting such evidence. Here, we provide constraints on sulfate deposition and diagenesis using geometric information provided by our newly developed technique to analyze X-ray diffraction generated by the Planetary Instrument of X-ray Lithochemisty (PIXL, (*10*)), one of the proximity science instruments on the arm of *Perseverance* (*4*).

At both study areas, the Shenandoah formation is composed of fine-grained, sulfate-cemented, planar-laminated and low-angle cross-stratified sandstone and siltstone intercalated with recessive and resistant, meter-scale beds preserving soft-sediment deformation structures (*6*). At Hogwallow Flats, the Shenandoah fm. dips 5 to 8 degrees to the southeast (*11*). Based on similar orientations and elevations, and shared sedimentological and lithological characteristics, the Hidden Harbor outcrop (Yori Pass member) is likely the lateral equivalent of the Wildcat Ridge outcrop (Hogwallow Flats member) (*6*). SuperCam (SCAM) detected sulfate at the Wildcat

Ridge and Hidden Harbor outcrops in visible/infrared (VISIR) and Raman spectra. The absence of hydration bands between 3200-3700 cm$^{-1}$ in the Raman spectra indicates the presence of anhydrite (*12*).

Abrasion patches (*8*), informally named Berry Hollow (BH) and Uganik Island (UI), were created by *Perseverance* on mission sol 504 and 612 at the Hogwallow Flats and Yori Pass members, respectively (**Fig 1**). SHERLOC (Scanning Habitable Environments with Raman and Luminescence for Organics and Chemicals) analyses of the light-toned veins and vugs at the Berry Hollow and Uganik Island abrasion patches confirm the presence of anhydrite with some hydrated phases also detected, likely hydrated Mg sulfates (*13*). These regions and the surrounding host rock were targeted by PIXL (*10*) on sols 505 (BH1) and 507 (BH2), and 614 (UI1) and 617 (UI2) (**Fig1C–F**). In these scans, PIXL generated quantitative elemental maps (*10*), and also detected diffraction peaks which can be used to analyze crystal textures (*2, 14*) and discern between a limited set of stoichiometrically similar minerals (*15*). Crystallographic texture provides information about the state of stress, pore fluid pressure, and growth mechanism of veins (*16*).

## Results
### Composition and mineral phases of the vein and vug cements

PIXL's X-ray fluorescence analysis of the scanned regions confirms that the light-toned material (visible in BH2, UI1 and UI2) is mainly composed of calcium and sulfur (reported as CaO + SO$_3$ here) and is largely devoid of other measurable elements (e.g., <10 wt% total other measured elements for the larger feature at UI1; **Figs 2, S2-4, S6**). The BH1 scan does not contain any CaO + SO$_3$ regions and is excluded from further analysis (**Fig S5**). The host rock is rich in FeO$_T$, MgO and SiO$_2$, but is largely free of CaO and depleted in SO$_3$ compared to the CaO+SO$_3$-rich regions (See **Figs S2-4**). Diffraction from the light-toned material in both scans was compared to a database (*17, 18*) of possible diffraction spectra detected by PIXL for all orientations of anhydrite (*19*), gypsum (*20*), and bassanite (*21*). Indexing was successful for 80% of all beam locations, with the remaining 20% being statistically ambiguous and therefore not included (see Supplementary Materials).

Exposing hydrated calcium sulfate (e.g., gypsum) to the Martian atmosphere may result in conversion to less hydrated states (e.g., bassanite). The analysis of sieved (< 150 µm) sulfates at Gale crater shows that minimal conversion is likely to occur after 4 sols while significant conversion is observed after 8 sols (*22*). MEDA observations suggest that the daytime environmental conditions give rise to the possibility that CaSO4 could be dehydrated during the day and rehydrated at night at temperatures below 195 K (*23*). However, prior to abrasion, sulfates remain stable in several different hydration states due to the process's slow kinetics and the overlying rock's protection. From the time of the abrasion, BH1 and BH2 were measured within 1 and 3 sols respectively, and UI1 and UI2 were measured within 2 and 5 sols. The proportions of different Ca-sulfate dehydration phases in BH2 and UI1 are approximately 37% and 20% gypsum, 45% and 66% anhydrite, and 18% and 14% bassanite, respectively. UI2 contains similar proportions of calcium sulfates as UI1 (**Fig. S7**), implying that no conversion had occurred between abrasion and data collection. Because many bassanite diffraction peaks overlap with those of gypsum and anhydrite, bassanite identification is unreliable (*22*), and these beam locations were removed from further analysis. Beam locations identified as gypsum and anhydrite (**Fig 2E, F**) show no clear spatial grouping. Proportions of gypsum relative to anhydrite at both locations imply that sulfates at Berry Hollow were deposited in a less saline and/or lower temperature (*24-26*) environment than those at Uganik Island.

### Outcrop-scale fracture network

The outcrop-scale fracture network at Hidden Harbor (**Fig. 1C, 3**) and Wildcat Ridge (**Fig. 1E, 3**) is well connected and polygonal, with a mean fracture spacing of 1.4 cm and 2.2 cm respectively, measured with the intercept method (*27*). The orientation frequency distribution of all fracture traces mapped within the horizontal outcrop plane displays two weak, mutually perpendicular modes of the orientation distribution at 295° and 205° (Hidden Harbor: **Fig. 1C, 3A**) 153° and 253° (Wildcat Ridge: **Fig. 1E, 3B**). Oblique outcrop photographs show boulders and steep edges of the rocky outcrop with mainly vertical fracture traces and some bedding-parallel fractures. Hence, most fractures are likely subvertical. In addition, there is no evidence for shear displacement along fractures. Therefore, the outcrop-scale fracture system at both outcrops likely represents subvertical extension fractures (*28*).

The frequency distribution of the orientation of exposed vein segments at Hidden Harbor resembles that of the entire fracture network, i.e., fractures with and without visible mineral infill, but with a much more pronounced mode at 295°. At Wildcat Ridge, sulfates mainly occur in eight separate decimeter-scale veins (**Fig. 3B**). As a result, the orientation frequency distribution of exposed vein segments is much less isotropic than that of the wider fracture network, with the strongest mode at 163° and two smaller modes at 323° and 303°. Both orientation frequency distributions (i.e., of the entire fracture network and the visible veins only) at Hidden Harbor show a marked minimum for ~NS-striking fracture traces. By estimating the orientation of the minimum principal stress within the outcrop plane ($\sigma_{Hmin}$) that maximizes the dilation tendency on the entire fracture network (*29*), we derive an $\sigma_{Hmin}$ trending ~10°/190° (NS, **Fig. 3**) at Hidden Harbor, supported by the relative scarcity of N-S-striking fracture traces. Similarly, maximizing the dilation tendency on the entire fracture network at Wildcat Ridge in the horizontal plane gives a trend of ~ 45°/225° for $\sigma_{Hmin}$ (NE/SW, **Fig. 3**), which is approximately orthogonal to the most common strike of the exposed vein segments.

The polygonal fracture networks formed in a stress field where the maximum principal stress $\sigma_1$ was vertical, most likely imposed by the overburden of the sediment or sedimentary rock present at the time of fracture formation (*30*), similar to that derived from 3D exposures of sulfate veins at Gale crater (*31*). However, in contrast to the fractures observed at Gale crater, the presence of a preferred orientation of exposed vein segments suggests that a small difference in magnitude of the horizontal stresses was present (intermediate and minimum principal stresses $\sigma_2$ and $\sigma_3$, respectively).

**Uganik Island abrasion patch veins**
The sulfate veins in the Uganik Island abrasion patch at Hidden Harbor (**Figs. 1D, 3A**) have lengths between 0.4 and 8 mm and widths between 0.1 and 4 mm, as measured by Feret's minimum and maximum diameter (*32*). Their aspect ratios range from 1.2 to 8.7 with a mean of 3. The largest sulfate regions have low aspect ratios and approximately rectangular shape. Several larger cracks display the shape of three-armed stars (**Fig 4A**). The veins and vugs appear largely isolated at the scale of observation. Omitting the low-aspect-ratio vugs, the orientation frequency distribution of traces shows two modes at right angles to each other, namely 240° and 150° (**Fig 4A**), yielding $\sigma_{Hmin}$ trending ~10°/190° (NS), in line with the outcrop analysis (**Fig 3A**).

The cement of the ~ 4 mm scale "fish-shaped" vug (*7*) was scanned with PIXL in the UI1 scan (**Figs. 2A, 3A**). The optical images reveal a blocky cement texture in this vug (**Figs. 2A, 3A**; (*7*)). The pole figures for the [010] reflections of gypsum and anhydrite do not show a crystallographic preferred orientation for the overall texture pattern (*33*) (**Fig 4B, C**). However, mapping the local order of the [010] pole (**Fig 4D**) for gypsum and anhydrite shows discrete regions of high local order (see Supplementary Materials), which are interpreted as individual crystals. This blocky crystal texture is not diagnostic of the fracture mechanism or fracture kinematics (*16*) and can be explained either by a fast nucleation rate during the growth of a primary vug cement in an open fluid-filled cavity or due to later static recrystallisation or replacement of an earlier cement (*16*). Both optical inspection and the size of domains with local order indicate a crystal size on the order of several hundreds of micrometers.

### Berry Hollow abrasion patch vein

The Berry Hollow abrasion patch at Wildcat Ridge exposes one sulfate vein (**Figs. 1F, 2B**) with the shape of a classical wing crack (*34, 35*). The vein has a straight ~ 4 mm long NS-striking central segment flanked by ~ 1 mm long curved, tapering termini, making a ~ ±35° angle with the central segment (**Fig. 4A**). The wing-crack geometry implies that the vein propagated as mixed-mode crack with a left-lateral shear component in the outcrop plane. The orientation of the maximum and minimum principal stresses ($\sigma_{Hmax}$ and $\sigma_{Hmin}$, respectively) within the horizontal outcrop plane can be determined from the wing-crack geometry (*35*) (**Fig. 2B, 5A**), giving a trend of ~ 45°/225° for $\sigma_{Hmin}$ (NE/SW), the same as the outcrop-scale fracture network (**Fig. 3**). Crystal-texture analysis demonstrates that both anhydrite and gypsum crystals of this vein exhibit a crystallographic preferred orientation. Pole figures of individual beam spots containing anhydrite (squares; space group Cmcm) and gypsum (circles; space group C2/c) show that most [010] planes of both minerals strike approximately parallel to the outcrop trace of the wall of the main vein segment, and their poles make an angle of about 45° and 40° to the inferred $\sigma_{Hmax}$ direction, respectively (**Fig 5, (*33*)**). The [010] plane of both minerals is inclined (**Fig. 5B**). Analysis of the fluorescence signal shows that the vein dips to the W at approximately 40° (**Fig. S8**). The vein geometry, orientation, and crystal texture imply that the central vein segment experienced both an opening and shear displacement with either an elongate-blocky or stretching-vein fabric (*16*). Therefore, our observations suggest that the vein at Berry Hollow probably formed as a mixed-mode crack-seal vein filled with primary Ca-sulfates.

Considering the parabolic nature of the brittle yield envelope for rocks in the tensile domain, this also means that this vein formed at greater differential stress (e.g. Figure 3 in Ref. (*16*)), and thus greater burial depth, compared to the extension fractures observed at Wildcat Ridge and Hidden Harbor. It is possible that some of the larger fractures mapped at the surface of Wildcat Ridge (**Fig 3B**) are also mixed-mode fractures with a dip angle < 90° and may have formed in a more consolidated rock at a greater burial depth than the vertical extension cracks observed elsewhere.

### Origin of the fracture network and state of stress

The outcrop-scale polygonal vertical extension fractures could have formed due to volumetric shrinkage of sediment at or near the surface (desiccation or syneresis, e.g. (*36, 37*)), or by hydraulic fracturing in partially or fully lithified sediment with over-pressured pore-fluid (*30*). While there is no observable diagnostic feature that allows us to ascertain which of these mechanisms operated, rock mechanics theory suggests that the maximum burial depth is constrained to a differential stress < 4T where T is the tensile strength of the rock (*28*). However, Berry Hollow exposes a moderately dipping mixed-mode crack-seal vein, which formed at a differential stress > 4T (*28*), and thus at greater burial depth than the polygonal, vertical extension fractures. For a rock with relatively low tensile strength (*38*), we obtain a minimum burial depth of ~ 80m during the formation of the Berry Hollow wing crack (**Fig. S9**).

## Discussion
### Sulfate diagenetic history

Analysis of the Ca-sulfate-filled fractures at the outcrops at Hidden Harbor and Wildcat Ridge and the Uganik Island abrasion reveals that they are vertical extension fractures that could have formed due to volumetric shrinkage of sediment at or near the surface (desiccation or syneresis, e.g. (*36, 37*)), or by hydraulic fracturing in partially or fully lithified sediment with over-pressured pore-fluid (*30*) at depths as shallow as 10s of meters (**Fig. 6B**). While the analysis does not indicate if the observed Ca-sulfates were a primary or replacement cement, thermodynamic constraints suggest that the original mineral(s) occupying these fractures may have included chloride minerals dominated by Na, K, Mg, and/or Ca, especially as $SO_4$ and dissolved inorganic carbon were progressively removed from the fluid by precipitation of Fe/Mg-carbonate- and Fe/Mg-sulfate minerals (*39*), the latter of which are abundant outside of the Ca-sulfate filled regions in the Uganik Island abrasion. During diagenesis, these fractures could then have been preferentially replaced by Ca-sulfates as late-stage and highly soluble phases were dissolved and replaced by relatively insoluble gypsum and/or anhydrite.

In contrast, analysis of the crack seal vein at the Berry Hollow abrasion formed at a minimum burial depth of ~ 80m, with Ca-sulfate as a primary cement (**Figs. 6C, S9**). Assuming the sedimentary fan front once extended at least 4km further E, and at least 3km further SE to Santa Cruz and Pilot Pinnacle buttes, respectively (*40*), and has since been eroded back to its current position, the 80 m minimum depth estimate can be accommodated by the ~110m elevation difference between Berry Hollow and the top of the sedimentary fan. Compared to the features analyzed at Uganik Island, the Berry Hollow Ca-sulfate has a higher proportion of gypsum. Differences in gypsum and anhydrite precipitation conditions can be related to variations in temperature and/or salinity (*24-26*), with gypsum favored at lower temperature and/or salinity. Therefore, since the gypsum-dominant crack-seal vein formed at greater depth, and therefore possibly greater temperature than the extension fractures elsewhere in the formation, it likely precipitated from a solution with lower salinity.

Despite the possibility of a replacive origin, Ca-sulfate in Kukaklek (from Hidden Harbor) contains abundant vein and vug cements likely formed in the shallow subsurface, while Ca-sulfate in the Hazeltop and Bearwallow samples (from Wildcat Ridge) preserve information from a warmer and less saline subsurface environment. These samples therefore represent two different paleoenvironments, capturing a range of saline surface- and groundwaters present at the Shenandoah formation providing two unique opportunities in the search for extraterrestrial biosignatures (*41-45*) on their potential return to Earth.

**Materials and Methods**

<u>PIXL data acquisition and processing</u>

X-ray fluorescence (XRF) data was collected by the Planetary Instrument for X-ray Lithochemistry (PIXL (*10*)). To create a map, PIXL is placed at a standoff distance of ca. 25.5mm from the surface and scanned via a hexapod, with data collected with a 10s dwell time in a regular grid with 120µm spacing. Each data acquisition point is given a PIXL Motor Control (PMC) number for each scan. A Rh X-ray tube operating at 28keV is used to generate X-rays with the incident beam focused to ca. 120 µm spot (at 8keV) at the surface using a polycapillary optic (*10*). Due to the nature of the polycapillary optic, the focus size depends on the X-ray energy resulting in a smaller beam size at higher energies and vice-versa (*14, 46*). Excited XRF photons are collected with a pair of silicon drift detectors (SSDs) placed on either side of the optic (*10*).

Individual XRF spectra for each detector and PMC were quantified as oxides with PIQUANT (*47*) and visualized in PIXLISE (*48*). Elemental quantifications for each PMC were cleaned to remove effects due surface topography and diffraction in PIXLISE (*49*). Data for $MgO$, $Al_2O_3$, $SiO_2$, $MnO$, $FeO_T$, $SO_3$, $CaO$ and $SrO$ were exported in units of mmol/g for the three-color images in **Fig. 2** and the ternary diagrams in **Fig. S6**, and wt% for the individual elemental images in **Figs. S2-5**. For BH1, BH2, and UI2, SrO values are below the limit of detection due to low signal and overlap with the Fe pileup peak and are therefore not presented.

## Mapping crystallography with PIXL

The process for mapping crystallography with PIXL is as follows. Firstly, XRF data was used to select PMCs that contained Ca-sulfate minerals. Secondly, the diffraction that the PIXL would observe from the identified possible Ca-sulfate minerals (gypsum, bassanite, and anhydrite) was modelled to create a look-up-table (LUT), in a process called the "forward simulation". Thirdly, the observed diffraction was compared to this LUT and the statistically most likely mineral was selected, along with its relative crystallographic orientation to PIXL. The following describes these individual processes in detail.

### *Forward simulation*

For each mineral, the angles ($\gamma$, $\delta$) of the diffraction vectors orthogonal to the diffracting planes, *g*, for each (hkl) that results in a d-spacing within the range 1keV < E < 15keV for a two-theta of 158 degrees were calculated (*14, 50*). Given PIXL's relative X-ray beam and detector locations, a subset of reflections will be recorded for a given instrument-crystal orientation. For a fixed crystal orientation, we define three angles to describe the relative position of PIXL: $\alpha$, $\phi$, and $\tau$. $\alpha$ and $\phi$ define polar and elevation angles of the incident X-ray beam relative to the crystallographic c-axis (*14*), while $\tau$ describes the rotation of the two PIXL detectors around the X-ray beam axis. Given PIXLs angular resolution of approximately ± 4 degrees (*14*), instrument-crystal orientations were simulated with an interval of 4 degrees with the rotation *t* reduced to the range 0 < *t* < 180 due to detector symmetry.

The relative diffraction peak intensities were calculated using Equation (1):

$$I(E) = A(E)|F_{khl}|^2 \left(\frac{1+cos^2(2\theta)}{sin^2(\theta)cos(\theta)}\right)_{hkl} \quad (1)$$

$$F_{hkl} = \sum_n f_n e^{2\pi i(hx+ky+lz)} \quad (2)$$

$$f_n = f_0 + f'(E) + if''(E) \quad (3)$$

where $F_{hkl}$ is given by Equation (2) with the energy-dependent atomic scattering factors given by Equation (3) and drawn from Chantler (2000) (*51*). *A(E)* is an additional scaling factor equal to the calculated intensity of the incident radiation (*47*) with the Rh L intensities replaced by theoretically determined values (*52*) (**Fig S10**). Once a reflection is recorded, the structure factor is defined for the specific diffraction angle, $\theta$, with the proportion of any partially detected diffraction intensities determined. The resulting diffracted intensities observed in the two detectors are then included in the forward simulation LUT for each of the three CaSO4 minerals: anhydrite (*19*), gypsum (*20*), and bassanite (*21*). Anhydrite was converted to space group Cmcm for consistency with previous literature (*33*).

### *Identifying Diffraction*

Diffraction was identified by comparing the two detector outputs for each beam location (*53, 54*), using diffraction identified by a machine learning algorithm (*14*) as a guide. Once the location of the diffraction was identified, a Gaussian peak was fitted to the difference in counts between the two detectors (detector A – detector B) with a width defined by the detector resolution and an amplitude in counts centered on the peak energy. A positive amplitude denotes a peak in detector A, while a negative amplitude denotes a peak in detector B. For each beam position the list of identified diffraction peaks was used to create a measured diffraction spectrum with intensities in counts. Photon shot noise, in the form of Poisson noise, was then added to these spectra before the spectra were normalized (**Fig. S11B, C**).

*Mineral identification*

The measured diffraction spectra, $\Psi(E)$, for each beam location in the $CaSO_4$ regions were compared to the simulated diffraction spectra in the LUT for each of the three minerals, $\Lambda(E)$, by calculating the normalized cross-correlation ($r$):

$$r = \frac{1}{n-1} \sum_E \frac{(\Psi(E) - \mu_\Psi)(\Lambda(E) - \mu_\Lambda)}{\sigma_\Psi \sigma_\Lambda} \qquad (4)$$

calculated over $n$ energy points, $E$, where $\mu$ and $\sigma$ are the mean and standard deviation, respectively. The values for $r$ were ranked and the simulated diffraction spectra with the highest ranked correlation selected. Due to crystal and instrument symmetries, multiple simulated spectra resulted in the same correlation in many cases.

The highest $r$ values for each mineral were compared by applying a Fisher transformation (*55*):

$$r' = atanh(r) \qquad (5)$$

and comparing the resulting normal distributions:

$$z = \frac{(r'_1 - r'_2)}{S} \qquad (6)$$

where:

$$S = \sqrt{\frac{1}{n_1 - 3} + \frac{1}{n_2 - 3}} \qquad (7)$$

with $n_1$ and $n_2$ being the number of observations in $r_1$ and $r_2$ respectively. One mineral was selected when the resulting p-value was $\leq 0.05$. A p-value $> 0.05$ indicates that a single mineral could not be reliably identified, with these beam locations omitted from further analysis. Furthermore, due to the difficulty phasing bassanite (*22*), beam locations that resulted in bassanite selection were also omitted (18% and 14% of beam locations for BH and UI, respectively).

*Crystallographic orientation and mapping local crystalline order*

The crystal orientation relative to PIXL can be determined from the simulated diffraction spectra through a the application of an orientation matrix on the three crystallographic axes where the angles $\alpha$, $\phi$, and $\tau$ (*14*) correspond to the Euler angles $\Phi$, $\varphi_2$, and $\varphi_1$ in the Bunge convention (*56*). Local crystalline order is calculated by comparing the $n$ determined crystallographic orientations, $A_i$, for a given PMC with the $m$ crystallographic directions, $B_j$, for the four nearest neighbors using the absolute value of the cosine similarity, $|SC|$, where $SC$ is defined as:

$$SC_{i,j}(A_i, B_j) = \frac{A_i \cdot B_j}{\|A_i\|\|B_j\|} \tag{8}$$

The cosine similarity, $SC$, returns unity for vectors pointing in the same direction, zero for perpendicular vectors, and negative unity for vectors that are pointing in opposite directions. Taking the absolute value of the cosine similarity removes the directionality of the alignment, and the output therefore ranges from zero for perpendicular alignment to one for parallel alignment. The local crystalline order and confidence are then calculated as the average and variance of $|SC|$, weighted by the inverse of the distance between the given PMC and its three nearest neighbors.

*Fracture analysis*

Outcrop-scale fracture traces at Berry Hollow and Uganik Island (**Fig. 3**) were hand-digitized as polyline segments and the resulting maps exported as vector files (in SVG format) for orientation analysis with the program FracPaq (*57*). Fractures in the abrasion patch at Uganik Island (**Fig. 4**) were hand-digitized as polygons. The resulting map was exported as binary raster image and skeletonized with ImageJ (*58*). The skeletonized binary raster image was also analyzed with FracPaq. The local strike direction of exposed vein segments within the outcrop-scale fracture networks was measured manually at the points shown in **Fig. 3** and used to create the related polar histograms. The horizontal principal-stress directions were determined by maximizing the dilation tendency on the entire outcrop-scale fracture network (*29*), using the orientations of the individual polyline segments of the digitized fracture network delivered by FracPaq.

**Acknowledgments**

This research was carried out in part at the Jet Propulsion Laboratory, California Institute of Technology, under a contract with the National Aeronautics and Space Administration (80NM0018D0004). We are grateful to the Mars 2020 team members who participated in tactical and strategic science operations. MWMJ would like to thank W. T. Elam, T. Wang, and L. D. Nothdurft for detailed discussions.

**Funding:**

Mars 2020 Participating Scientist Program grant #80NSSC21K0328 (SJV, ALK)

JPL Subcontract grant #1529702 (JAH)

NASA Mars 2020 PS grant # 80NSSC21K0331 (KLS)

MCIN/AEI/10.13039/501100011033/FEDER, UE. grant # PID2022-140180OB-C21 (M-PZ)

**Author Contributions:**

Conceptualization: MWMJ, DTF, JAH, MMT, CES

Data curation: MWMJ, SJV

Formal analysis: MWMJ, CES, DCC, SJV, ALK, BG, SD, PN

Funding acquisition: ACA

Investigation: MWMJ, CES, APB, BJO, KPS, AOS, JM-F

Methodology: MWMJ, CES, SJV, ALK, BG, CMH, BJO

Project administration: ACA

Resources: MWMJ, SJV, ALK, BG, CMH

Software: MWMJ, BJO

Supervision: DTF, JAH, MMT, ACA

Validation: MWMJ, CES, BJO

Visualization: MWMJ, CES, LAW, SD, PN, APW, AEG

Writing – original draft: MWMJ, DTF, JAH, MMT, CES, NJT, SJV, KLS, KCB, APB, M-PZ

Writing – review & editing: All authors reviewed and edited the manuscript

**Competing interests:** Authors declare that they have no competing interests.

**Data and materials availability:** Data is available in the NASA Planetary Data System https://pds.nasa.gov/ PIXL: doi:10.17189/1522645. Image data from the following cameras was used (see **Table S1** for specific image names): HiRISE images: doi:10.17189/1520303, Navcam (image prefix NLF) and Cachecam (image prefix CCF): doi:10.17189/1522847, SHERLOC Imaging (image prefix SI1, SC3, SIF): doi:10.17189/1522846, Mastcam-Z (image prefix ZLF, ZRF): doi:10.17189/1522843.

**Figures**

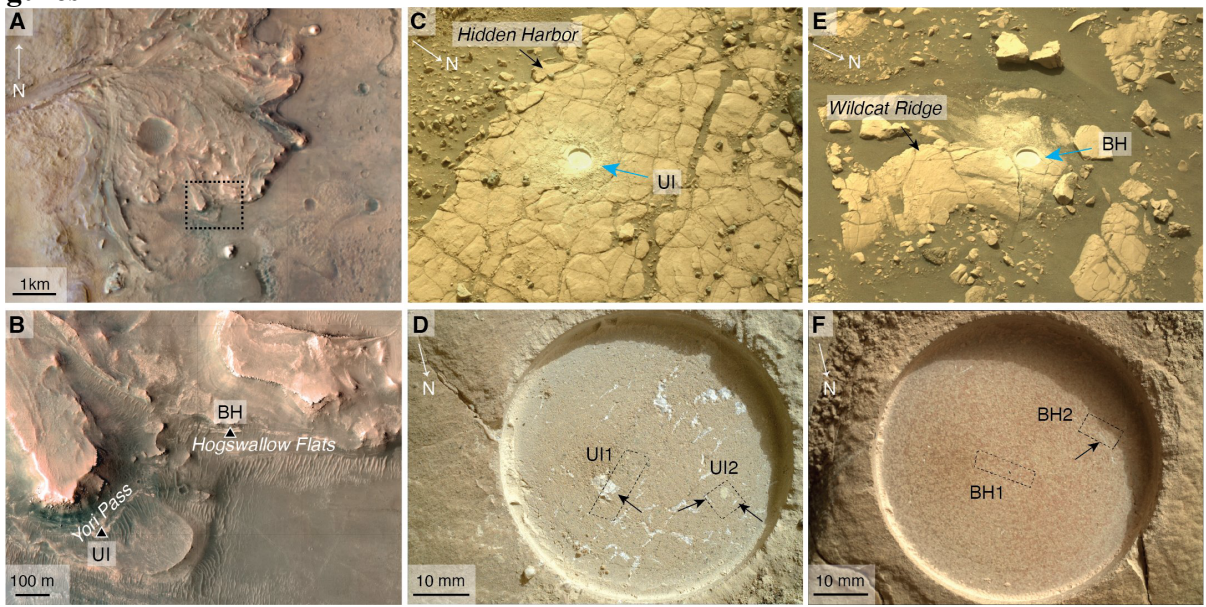

**Fig. 1. Location of the two abrasion patches at the Western Jezero fan** (**A**), (zoomed in: **B**) showing the locations of Uganik Island (UI) and the Berry Hollow (BH) abrasion patches. The location of each abrasion patch on each outcrop is shown in (**C, D**) and (**E, F**) for UI and BH respectively. Blue arrows in (C), (E) indicate the location of each patch and the outcrop-scale fracture network can be easily seen in each case. Dashed boxes in (D, E) show the scan footprints for UI1, UI2, BH1, and BH2, and black arrows and in (D, F) highlight $CaSO_4$ regions within the PIXL scan footprints. No substantial $CaSO_4$ features were observed by PIXL in BH1. Images in (C) and (E) were generated by Navcam, and (D) and (F) by WATSON.

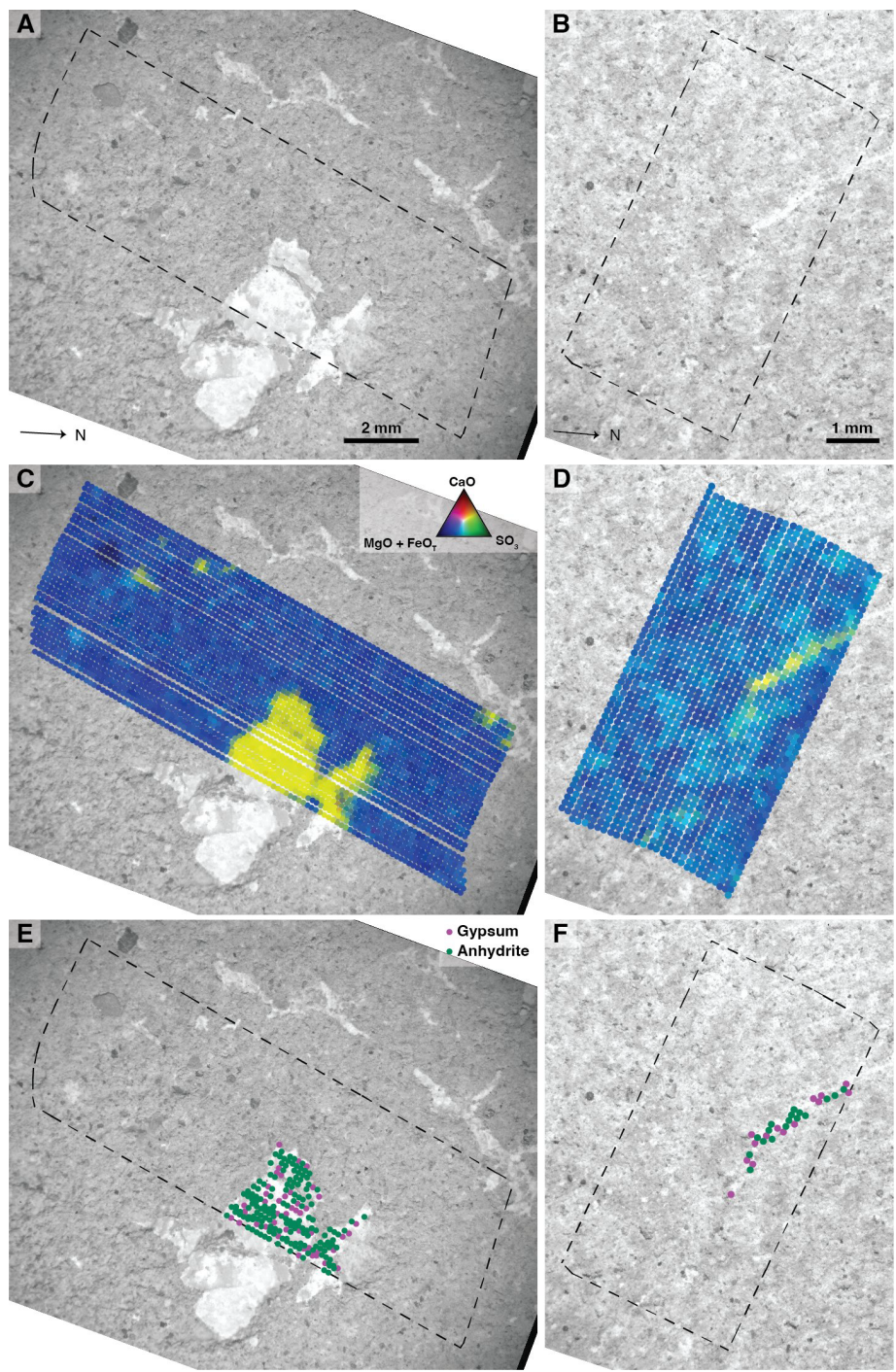

Fig. 2. PIXL analysis indicating Ca-sulfate rich regions in the two abrasion patches. Scan footprints for UI1 and BH2 (A, B) shown overlaid on SHERLOC ACI images. The scan locations are shown in Fig 1D and 1F, respectively. Light toned regions are interpreted as CaSO4, which is confirmed by mapping CaO (max 4.8, 7.9 mmolg$^{-1}$), SO$_3$ (max 6.0, 8.3 mmolg$^{-1}$), and FeO$_T$ + MgO (max 7.3, 7.7 mmolg$^{-1}$), as three-color RGB images (C, D). CaSO$_4$ minerals appear yellow according to the color mixing triangle in C (applies to C and D). CaSO$_4$ mineral identification maps for the two scans are shown in E and F show that anhydrite is the dominant phase.

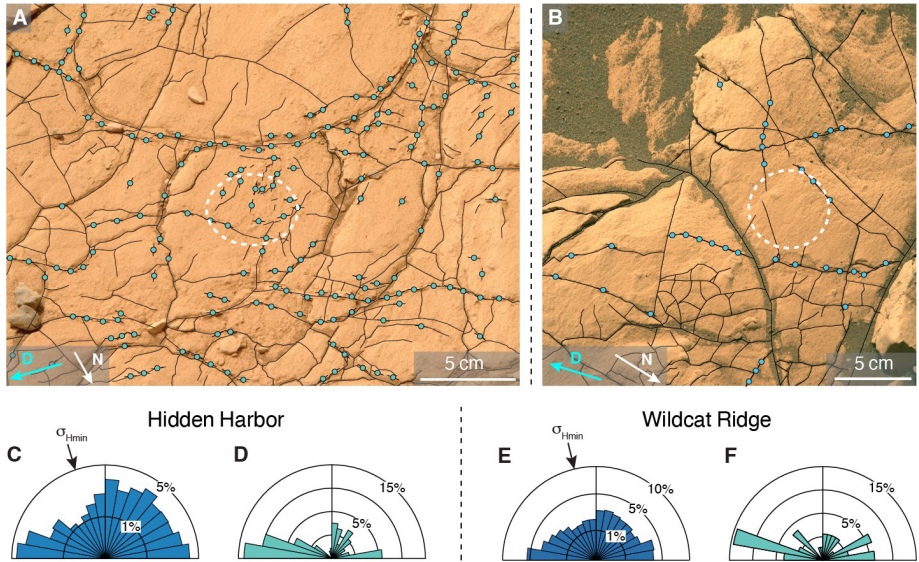

**Fig. 3. Fracture analysis of the studied outcrops.** Mastcam-Z photographs of the decimeter-scale fracture networks at (**A**) Hidden Harbor (HH) and (**B**) Wildcat Ridge (WR). Black solid lines mark fracture traces while the blue circles indicate locations where sulfate veins are exposed. The dashed white ellipses in (**A, B**) indicate the location of the Uganik Island and Berry Hollow abrasion patches respectively. (**C, D**) Relative frequency orientation distribution of fracture traces and sulfate vein segments at HH respectively visualized as polar histograms. Panels (**E**) and (**F**) show the equivalent data for WR. The black arrow at the histogram perimeter in (**C, E**) indicates the $\sigma_{Hmin}$ (*29*) orientation of ~10°/190° NS at HH and ~45°/225° NE/SW at WR. This direction aligns with the orientations derived from abrasion patch analyses (see **Figs. 4, 5**). White and cyan arrows in (**A**) and (**B**) show the direction of Martian North "N" the downhill direction "D" and apply to the relevant polar histograms.

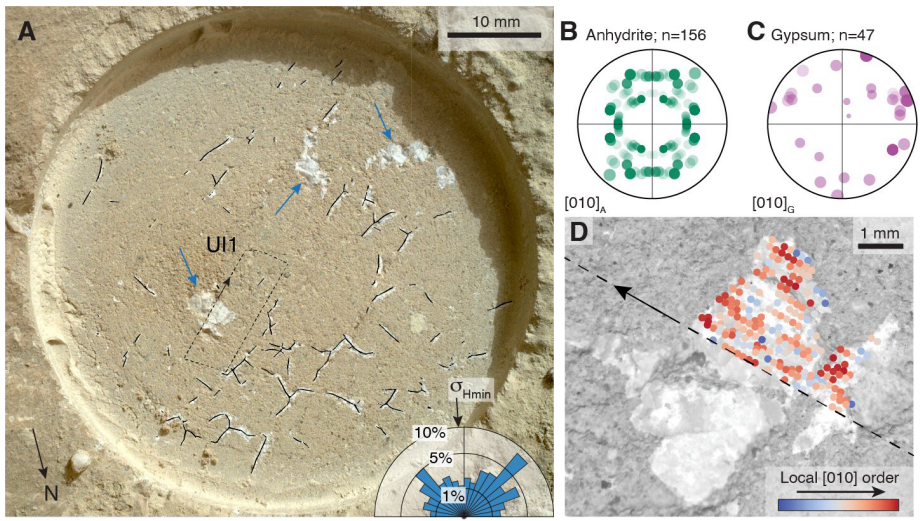

**Fig. 4. Uganik Island abrasion patch analysis.** SHERLOC WATSON image of the Uganik Island abrasion patch (**A**) showing sulfate-filled fractures (white regions) overlaid with a binary image of fracture traces. The blue arrows mark polygonal vugs with low aspect ratio omitted in the orientation analysis. Polar histogram of the orientation of fracture traces is shown in the lower right, indicating a $\sigma_{Hmin}$ direction ~10°/190° NS. Random [010] pole figure for anhydrite (**B**) and gypsum (**C**) at UI2 (dashed box in **A**) indicates no CPO. Mapping the local order of the [010]

plane for both anhydrite and gypsum (**D**) reveals regions of high local order throughout the area, suggesting a blocky texture. This blocky texture can also be seen in the sulfate mass immediately to the lower left of the mapped region. Black arrows in (**A**) and (**D**) show relative orientation between abrasion patch and scan.

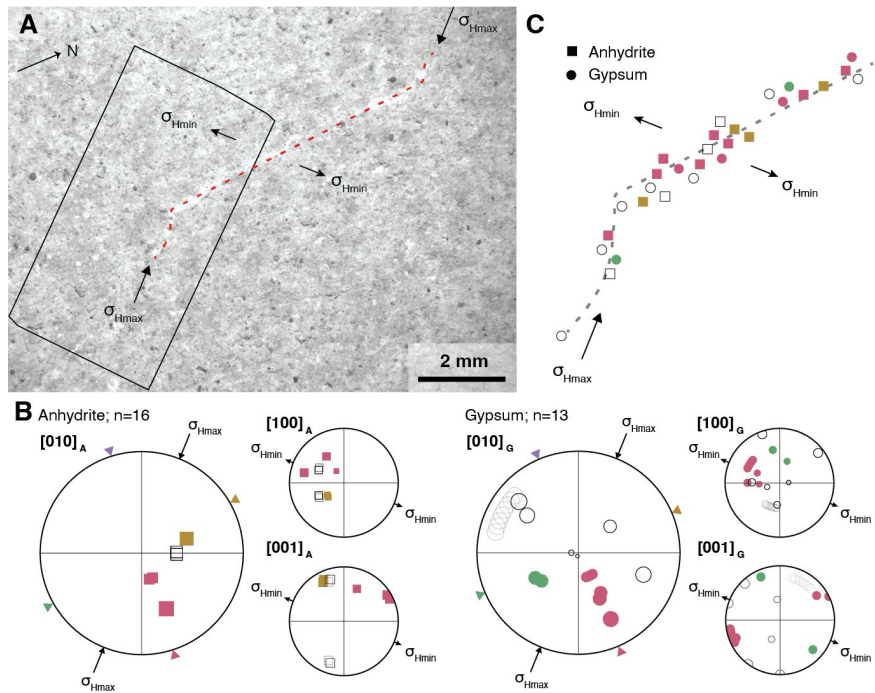

**Fig. 5. Berry Hollow abrasion patch analysis**. Visual inspection of the vein geometry at Berry Hollow (**A**, dashed line) permits identification of $\sigma_{Hmax}$ and $\sigma_{Hmin}$ directions with $\sigma_{Hmin} \sim 45°/225°$ (NE/SW). Beam locations identified as gypsum (circles) and anhydrite (squares) in **Fig 2E** are shown in stereographic projections with combined hemispheres in **B**. The colors of each beam location correspond to alignment of the [010] to the expected direction (shown as triangles on the outer perimeter of the stereo projections) for anhydrite (40° from $\sigma_{Hmax}$) and gypsum (45° from $\sigma_{Hmax}$) for the given principal stress direction $\sigma_{Hmax}$. The [100] and [001] figures are oriented in the expected orthogonal planes to the [010] figure and provide additional constraints on crystal orientations. Open squares and circles depict results from beam spots where [010] planes are not aligned with respect to the vein geometry. Mapping the beam locations (**C**) shows high alignment of the [010] plane throughout the vein, confirming the stress field identified through visual inspection.

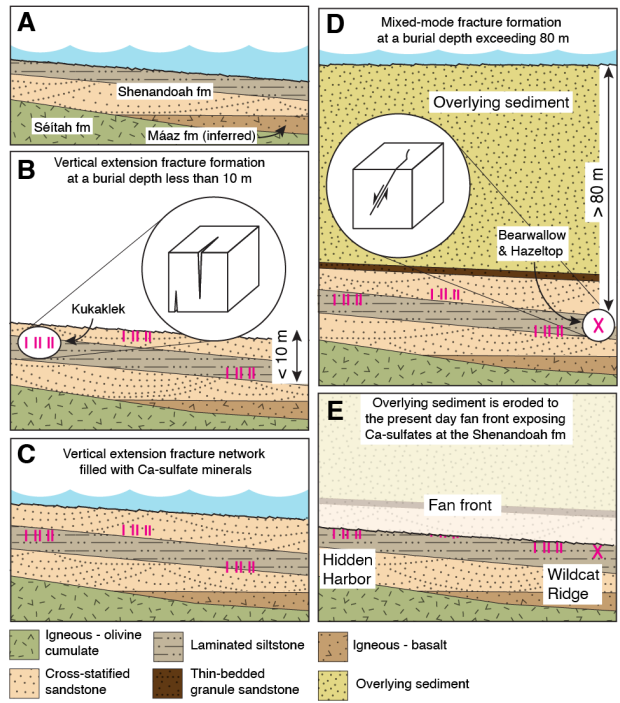

**Figure 6: Deposition sequence.** The sedimentary Shenandoah formation is deposited over the igneous Séítah and inferred Máaz formations (**A**) (*6*). Vertical extension fractures (vertical magenta lines) form in the shallow subsurface (**B**). These fractures are filled with primary or secondary Ca-sulfates (**C**). More than 80 m of additional material is deposited over the Shenandoah formation causing a mixed mode shear fracture (magenta X) to form (**D**) with primary Ca-sulfate precipitated. Excess material is eroded exposing a mixed-mode shear fracture at the Berry Hollow abrasion patch at Wildcat Ridge and vertical extension fractures throughout the formation (**E**).

**Supplementary Materials**

Significance of differences in fracture-wall roughness at Berry Hollow and Uganik Island

The fractures and veins at Wildcat Ridge (abrasion patch Berry Hollow) are generally narrower with straighter margins than those at Hidden Harbor. Since the fracture roughness of rocks generally decreases with decreasing grain size (*59*), this difference could be explained by textural differences in the host rock, which is finer-grained in Berry Hollow compared to Uganik Island (*7*). Alternatively, the degree of lithification at the time of fracture formation could have been different. Therefore, it is possible that the fracture network at Wildcat Ridge either formed at greater burial depth than that at Hidden Harbor or the difference is due to the finer grain size at Wildcat Ridge.

Causes for vertical maximum principal stress, the preferred orientation of veins, and the triaxial stress state

A vertically oriented maximum principal stress, as derived for the fracture networks at Berry Hollow and Uganik Island, is expected in a laterally constrained layer of rock under gravitational loading due to Poisson's effect (e.g., (*30*)) and does not require tectonic forces. Interestingly, the fracture traces, and much more so the vein traces, at both outcrops show a preferred orientation in the horizontal plane (**Fig. 3**). This preferential orientation of vein segments could be due to observational bias. The most frequently observed vein orientations are subparallel to the local topographic downhill direction (**Fig. 3**). Therefore, the mapped vein segments could have been exposed preferentially due to wind moving along the slope, biasing the documented orientation distribution. However, the veins could have also formed in a true triaxial state of stress ($\sigma 1 > \sigma 2 > \sigma 3$), even in the absence of tectonic forces. A deviation from perfect isotropy of fracture trace orientation is observed in experimental desiccation cracks formed in laterally unconstrained layers and can be easily produced by lateral changes in layer thickness, material properties (density, tensile strength, friction coefficient, pore-fluid pressure, moisture, etc.), basal friction with the underlying sedimentary beds, and geometry of the basal (or overlying) bed interface (*60*).

Further discussion of vein textures in the Berry Hollow wing crack

The geometry of the vein studied in detail at Berry Hollow resembles a wing crack (**Figs. 2, 5**), which implies that the central vein segment experienced both an opening and shear displacement. The crystal texture with its notable crystallographic preferred orientation (CPO) supports this interpretation. On Earth, syntaxial veins with elongate-blocky fabric and stretching veins have been observed to display a CPO (*16*). These vein fabrics are attributed to crack-seal veins formed under deviatoric stress (*16*). In addition, terrestrial gypsum and anhydrite commonly form mineral fibers in veins in antitaxial veins (*16, 61*), which can also exhibit a CPO. However, this CPO is weaker than that documented at Berry Hollow. In fibrous gypsum, also known as satin spar, the a-axes are aligned but with random rotations of the b- and c-axes around the a-axis (*62*). Fibrous anhydrite displays similar behavior with an alignment of the c-axes alone (*61*). Due to the lack of this rotational symmetry, we interpret the crystal texture of the vein at Berry Hollow as that of a crack-seal vein rather than an antitaxial vein. Interestingly, the [010] planes of gypsum, and those of anhydrite, strike approximately parallel to the strike expected from shear fractures formed under the stress state derived from the wing-crack geometry (**Fig. 5**). A similar CPO has been observed in deformation experiments on natural rehydrating anhydrite that is compacted under constant differential stress (*33*). Given the parabolic shape of the brittle yield envelope for rocks in the tensile domain, this also means that this vein formed at greater differential stress (e.g., Figure 3 in Ref. (*16*)), and thus greater burial depth compared to the vertical extension fractures observed elsewhere. Hence, it is also possible that some of the larger fractures mapped at the surface of Wildcat Ridge are also mixed-mode fractures which should have a dip angle < 90°.

Estimation of burial depth

The grain sizes at Berry Hollow and Uganik Island correspond to very fine to fine grained sand, and very fine to medium grained sand respectively (*7, 63*). The tensile strength ($T_s$) for fine grained sandstones varies, depending on the degree of cementation, and ranges from 75 kPa to 750 kPa for weakly cemented sandstones (*38*). For mixed-mode fractures such as the wing crack at the Berry Hollow abrasion, $\Delta\sigma > 4T_s$ (*28*). Therefore, for weakly cemented sandstones, as described above (*38*), we obtain minimum burial depths between 80 and 800 m with a Poisson's ratio for fine grained sandstone of ~0.29 (*64*) and a density of 1.68 gcm$^{-3}$ (*65*) (**Fig. S11**). Using geomorphological evidence and crater counting, it has been estimated that ~ 70m of material has been removed from above the crater floor (*66*). The outcrops studied here are within ~10 - 20 vertical meters of the crater floor contact (*6*). Our quantitative estimates of minimum burial depth are thus in close agreement with the lower bound estimated by these analyses.

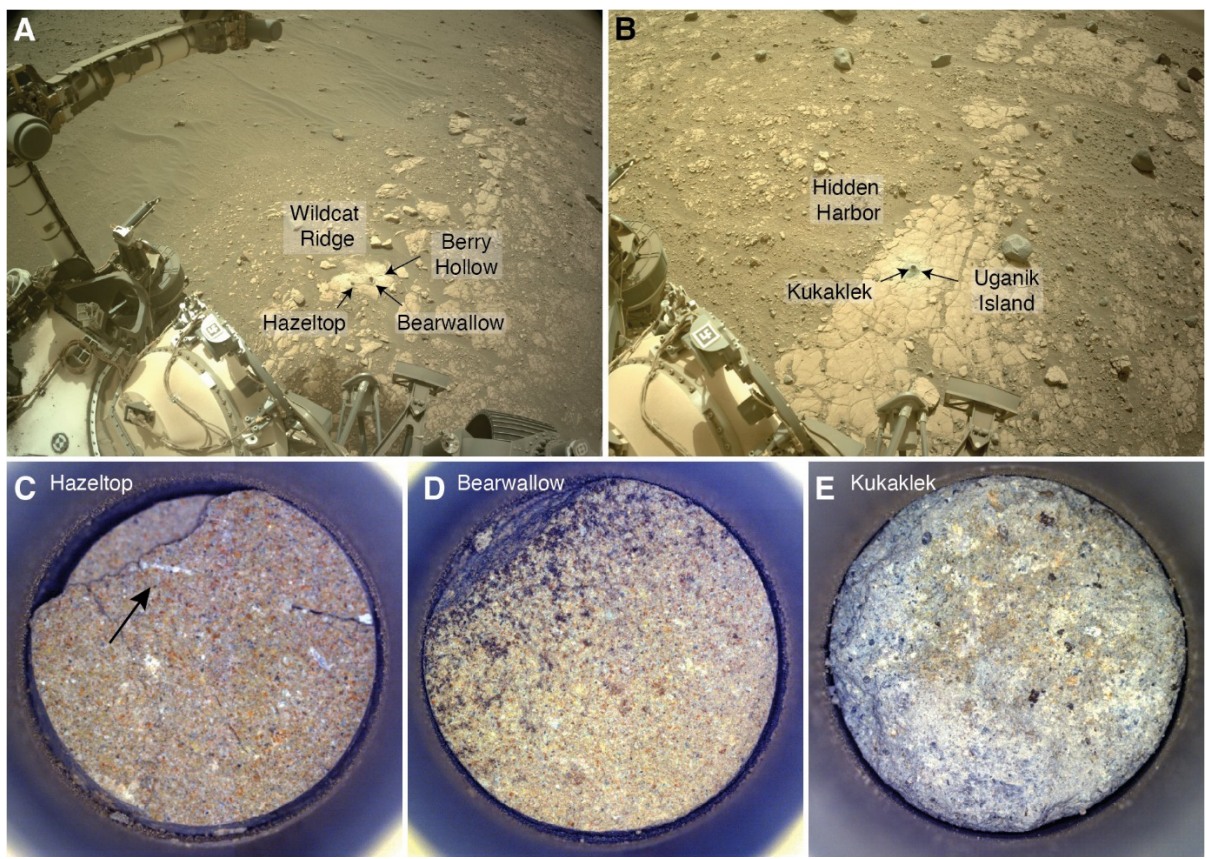

**Fig. S1.**
Location of the three cores for MSR at Wildcat Ridge (Hogwallow Flats member) (**A**) and Hidden Harbor (Yori Pass member) (**B**). Cachecam (*67*) (**C-E**) images of the Hazeltop (sealed on sol 509), Bearwallow (sealed on sol 516), and Kukaklek (sealed on sol 631) cores in the bit. The diameter of each rock core is 13 mm. A clear sulfate vein is visible in the Cachecam image of the Hazeltop core sample (black arrow, C), similar to that observed at Berry Hollow (**Figs. 1, 2, 5, S2**).

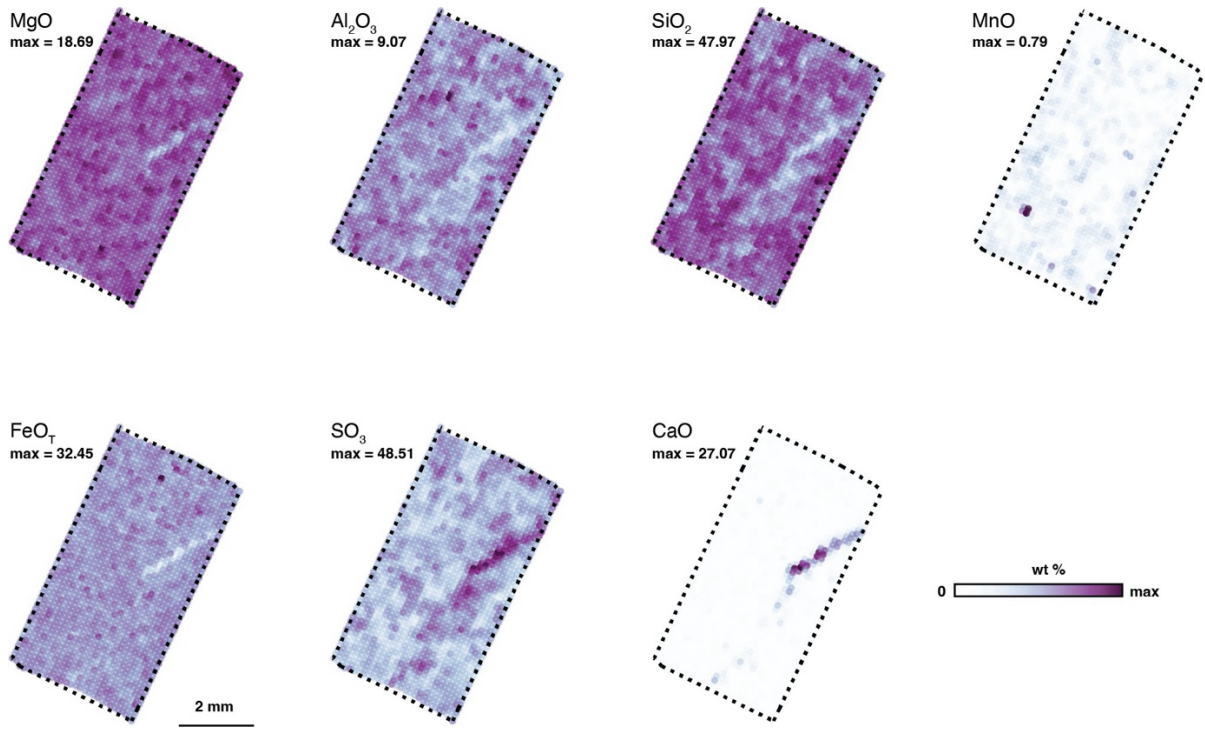

**Fig. S2.**

Elemental Maps for BH2. Concentration maximum is listed for each panel. All panels have the same zero point, with the colormap referring to all panels. The location of each scan is shown in **Fig 1D** and **Fig 2A**.

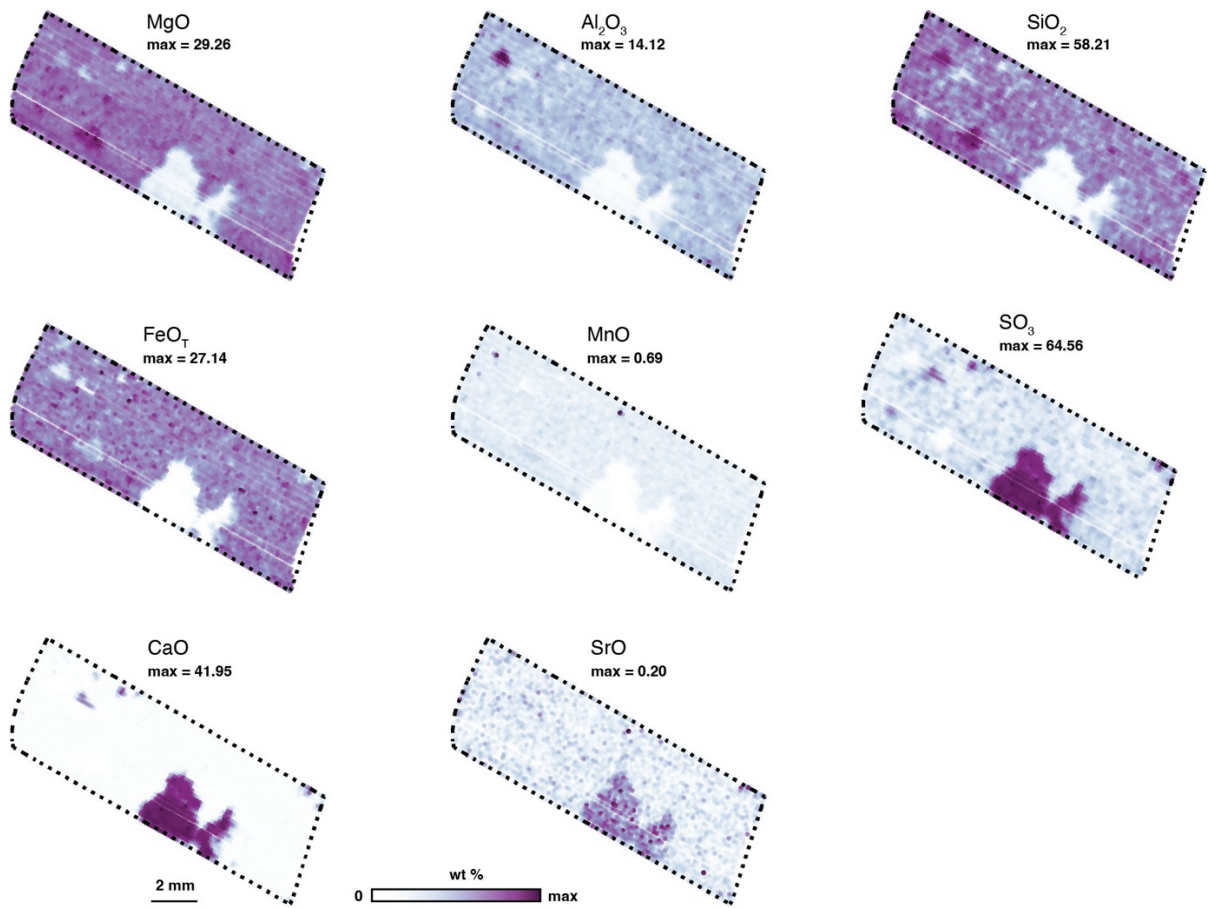

**Fig. S3.**
Elemental Maps for UI1. Concentration maximum is listed for each panel. All panels have the same zero point, with the colormap referring to all panels. The location of each scan is shown in **Fig 1F** and **Fig 2B**.

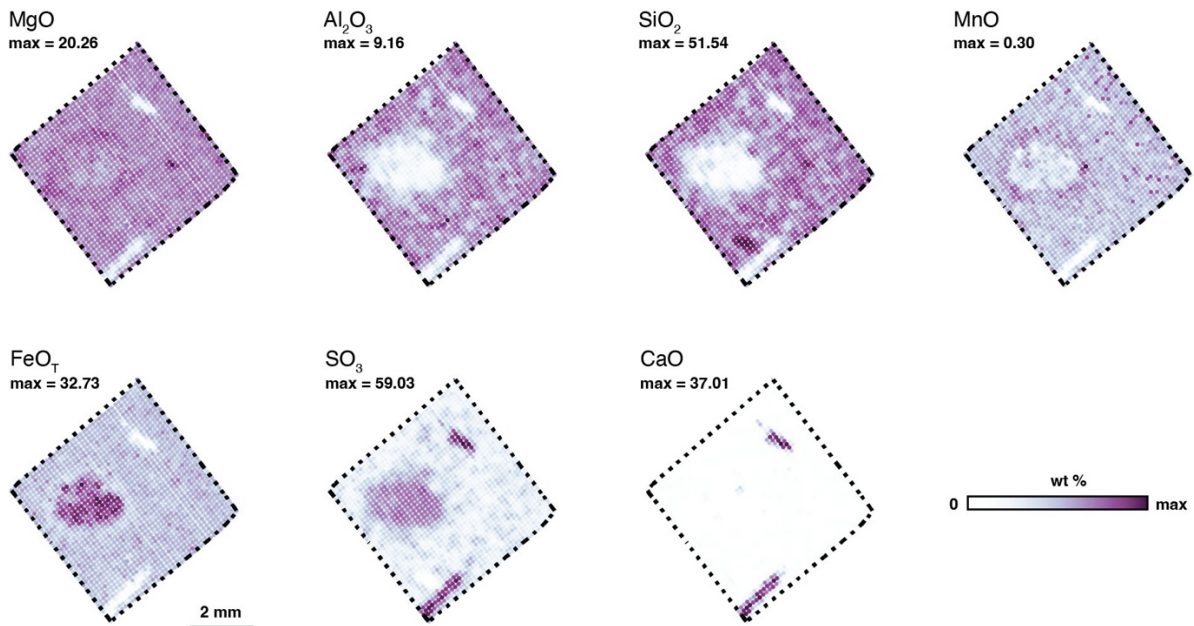

**Fig. S4.**
Elemental Maps for UI2. Concentration maximum is listed for each panel. All panels have the same zero point, with the colormap referring to all panels. The location of each scan is shown in **Fig 1F**.

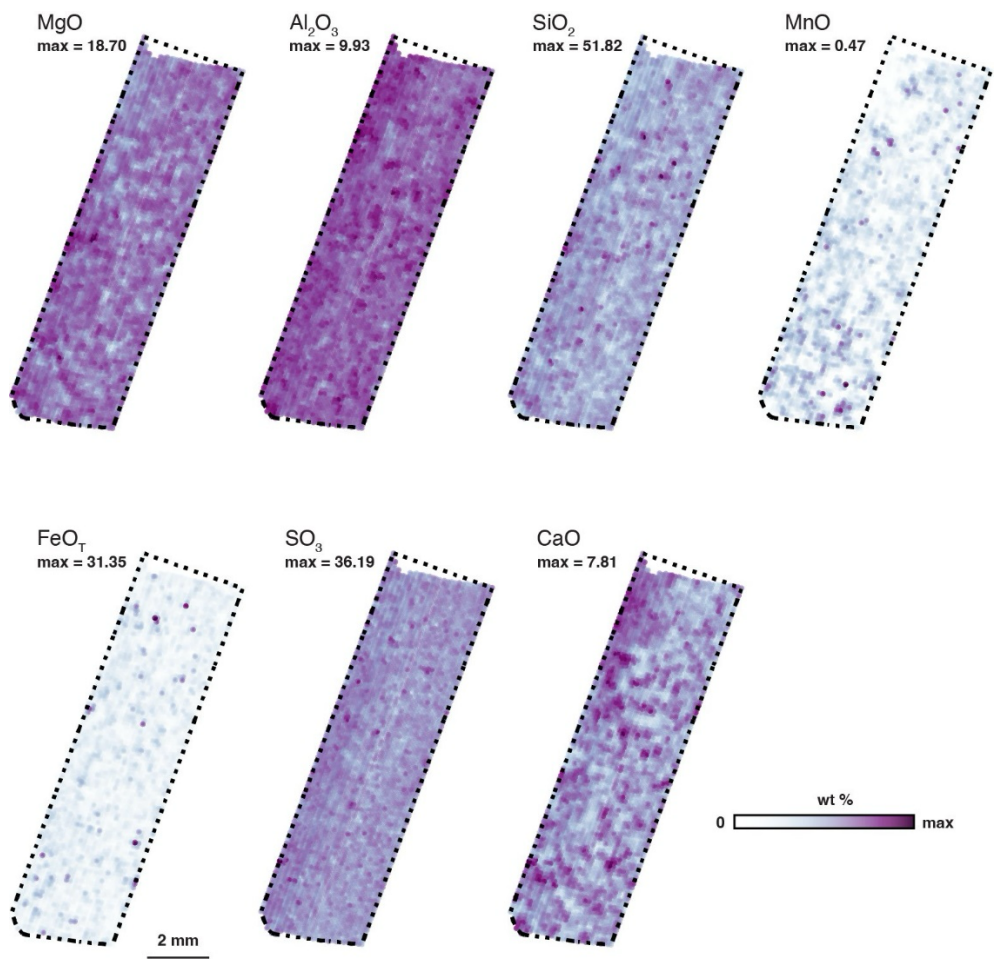

**Fig. S5.**
Elemental Maps for BH1. Concentration maximum is listed for each panel. All panels have the same zero point, with the colormap referring to all panels. The location of each scan is shown in **Fig 1D**.

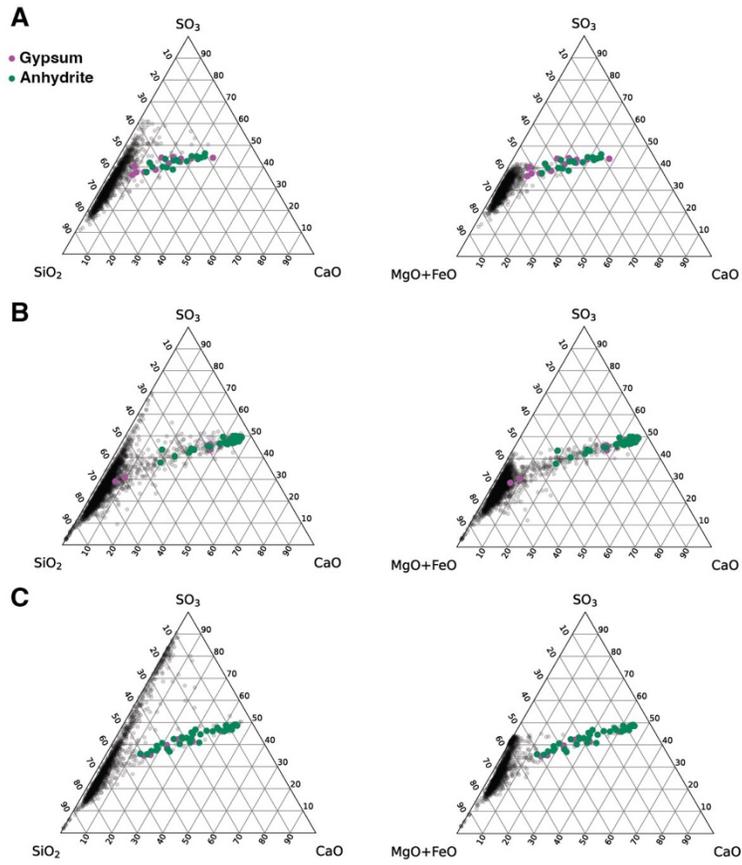

**Fig. S6.**
BH2 (**A**), UI1 (**B**), and UI2 (**C**). Gypsum and anhydrite identified in **Fig 2** (BH2 and UI1) and **Fig S7** (UI2) are in purple and green respectively.

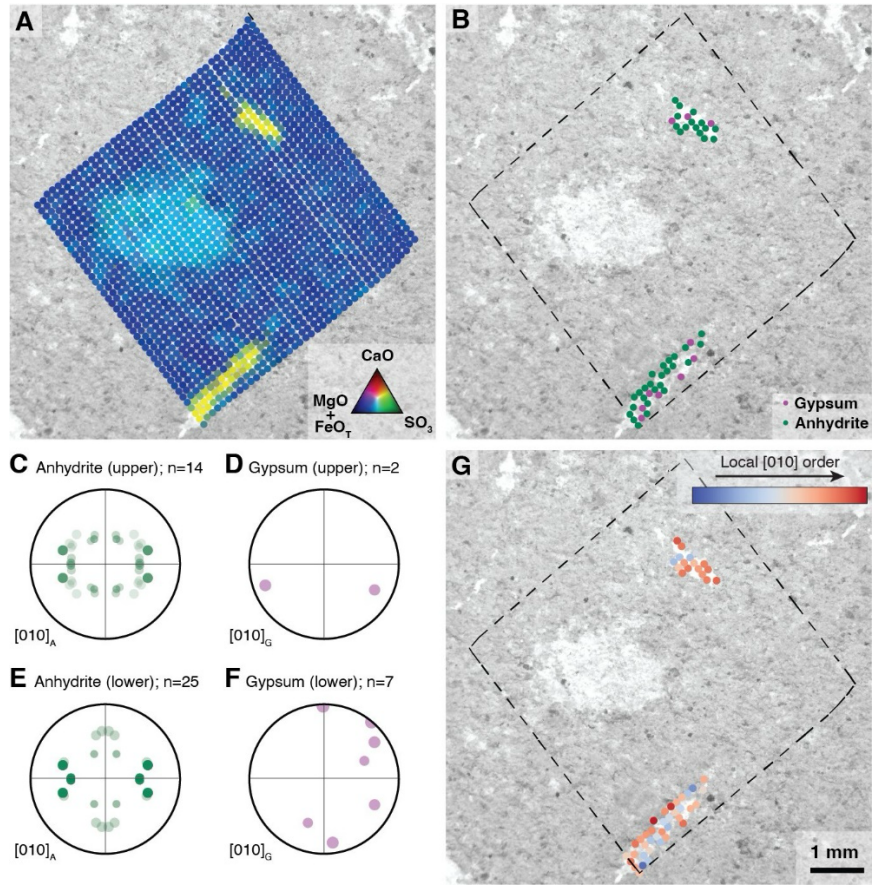

**Fig. S7.**

PIXL results for UI2 shown overlaid on a SHERLOC ACI image. The scan location is shown in **Fig. 1F**. Three color RGB image of CaO (max 4.8, 7.9 mmolg$^{-1}$), SO$_3$ (max 6.0, 8.3 mmolg$^{-1}$), and FeO$_T$ + MgO (max 7.3, 7.7 mmolg$^{-1}$) (**A**). CaSO$_4$ minerals appear yellow according to the color mixing triangle in (A) confirms that CaSO$_4$ minerals are present in the light toned regions. Mineral identification map (**B**) shows that anhydrite is the dominant phase. The mineral proportions are 70% anhydrite, 18% gypsum, 12% bassanite. As discussed elsewhere, locations identified as bassanite were discarded from further analysis. Random [010] pole figures for both upper (**C**, **D**) and lower (**E**, **F**) features over the region of interest indicates no CPO in either case. Mapping the local order of the [010] plane for both anhydrite and gypsum (**G**) reveals regions of high local order throughout the areas, suggesting a crystalline blocky texture. The scale bar in (**G**) also applies to (**A**) and (**B**).

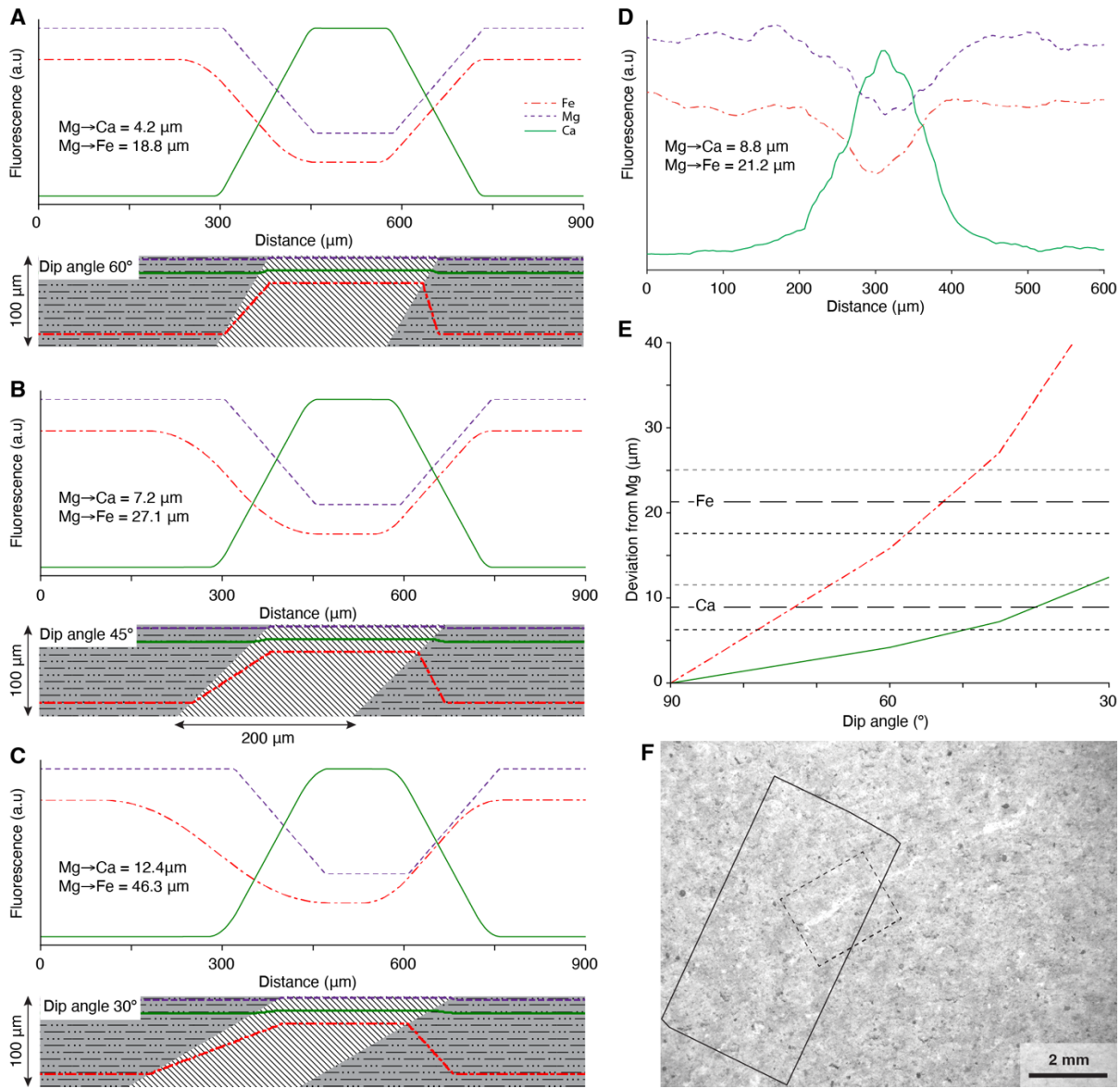

**Fig. S8.**
Simulations of the effect of vein dip angles on the fluorescent signal for a calcium sulfate (modelled as $CaSO_4$, $\rho = 2.32$ gcm$^{-3}$) vein with a surface exposure of 200 μm in a Fe-Mg rich silica host rock (modelled as $SiO_2$ with Gale crater bedrock density, $\rho = 1.68$ gcm$^{-3}$ (*65*)). Absorption coefficients from Elam *et al.*, 2002 (*68*). The profile of the expected fluorescence yield is dependent on the maximum escape depth of the fluorescent photons (lines plotted on schematics) and the dip angle of the vein. With a shallow escape depth, Mg accurately measures the vein exposed at the surface. As the escape depth increases the fluorescent signal maps more of the subsurface vein, therefore shifting towards the direction of the vein dip. For the case of a host rock density of 1.68 gcm$^{-3}$ and a vein dip of 30° to the surface (**A**), Ca and Fe are shifted 9.7 and 33.9 μm relative to Mg, respectively. For the case of a vein dip of 45° and 60° to the surface (**B, C**), Ca and Fe are shifted 3.2, 5.6 and 11.4, 19.7 μm respectively. For a vertical vein with a dip of 90° to the surface, all there is no shift between the various elemental profiles. Analysis of the vein at BH (**D**) shows that Ca and Fe are shifted 8.8 and 21.2 μm relative to Mg respectively. Plots of the deviation from Mg as a function of dip angle for Fe and Ca with a host rock density of 1.68 gcm$^{-3}$ (**E**). The dashed horizontal lines indicate the fitted deviation from Mg for Fe and Ca (**D**), with the dotted lines indicating the 95% confidence bounds of the fitted peak position. Comparing fitted profiles for the vein segment at BH2 (**D**) and the simulated data (**E**), we find a dip angle relative to the surface of ~ 50°. (**F**) shows the location of the profiles in (**D**) (dashed box) and the extent of the scan area (solid box).

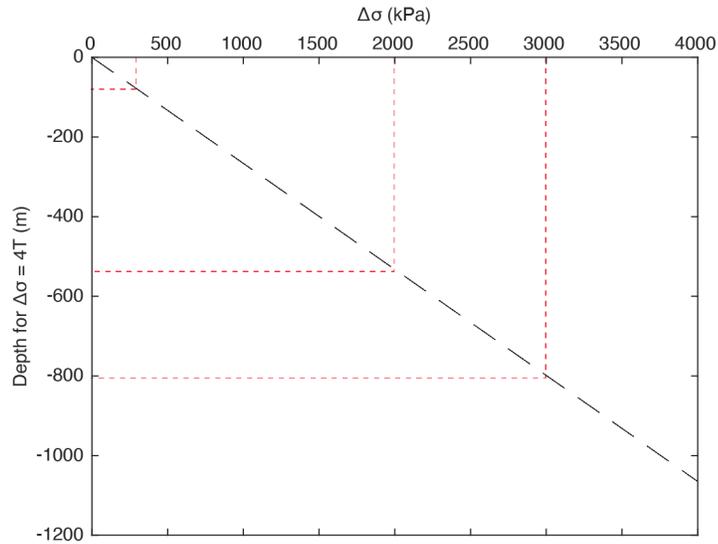

**Fig. S9.**
Relationship between the difference of the vertical and horizontal stress components ($\Delta\sigma$) and the burial depth using the following parameters: rock density $\rho = 1.68$ gcm$^{-3}$ (*65*), gravitational acceleration $g = 3.73$ ms$^{-2}$, Poisson's ratio $n = 0.29$ (for fine grained sediment (*64*)). Letting $h$ denote burial depth, $\Delta\sigma$ is calculated as (*28*): $\Delta\sigma = \rho g h \left(1 - \frac{1}{(n^{-1}-1)}\right)$. For tensile strengths (*T*) of 75 kPa, 500 kPa and 750 kPa, ($\Delta\sigma = 4T = 300$ kPa, 2 MPa and 3 MPa) (*38*) we obtain burial depths of 80 m, ~530m and 800 m, respectively (dashed red lines).

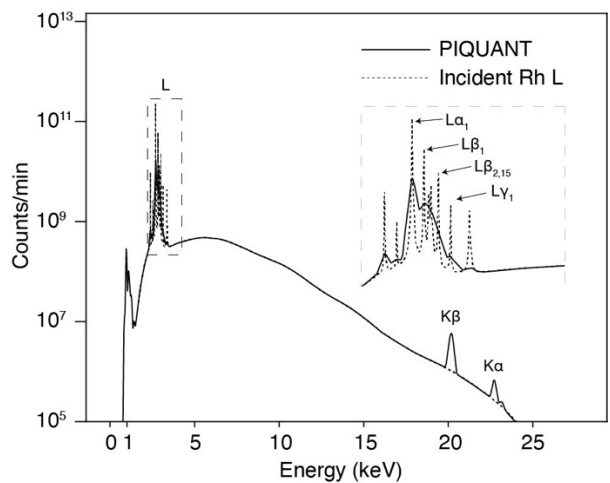

**Fig. S10.**
Incident intensity spectrum is most intense around the Rh L emission lines, with a highly non uniform intensity. This leads to relative increased intensities of some diffraction peaks, notably anhydrite (022) excited by the L$\alpha_1$ peak at 2696.8 eV (inset).

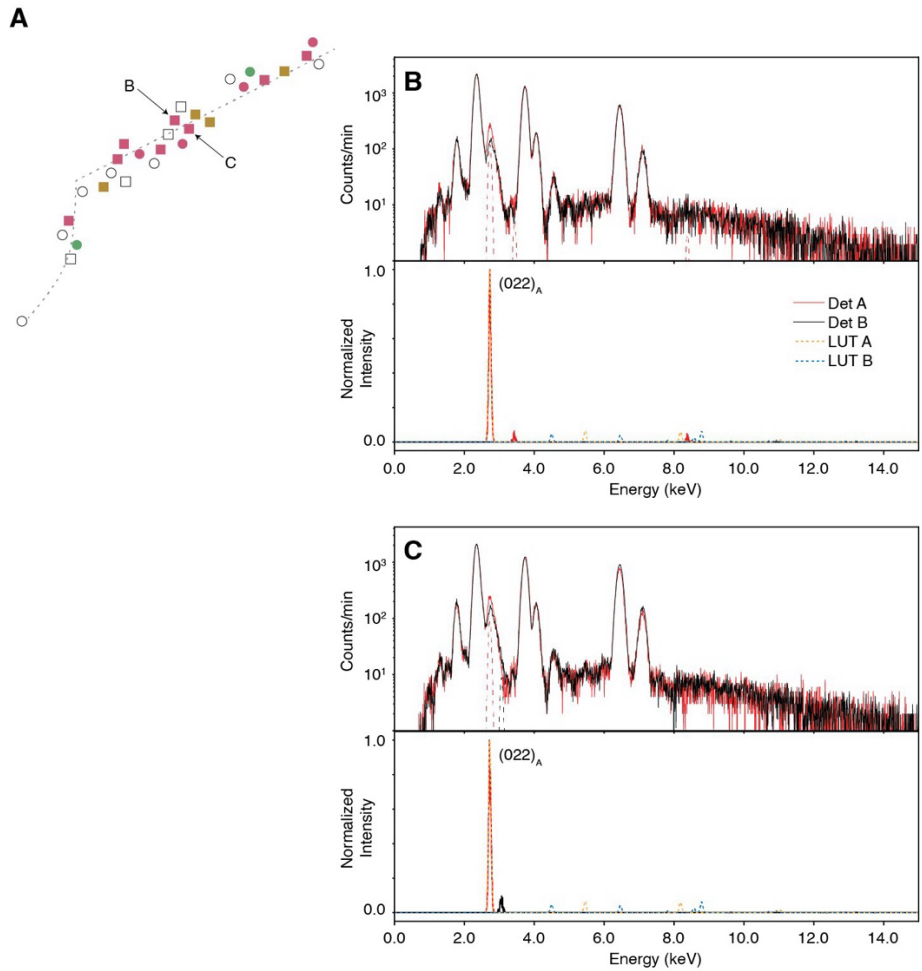

**Fig. S11.**
Two spatially adjacent beam locations identified as anhydrite with similar crystallographic alignment (**A**). Panel A is described in **Fig 4C**. The prominent peak in the data (red dashed peak in upper plots of **B**, **C**) has an excellent correlation (cross-correlation coefficient = 0.98 in each case) to the anhydrite (022) peak from the look up table (LUT, orange dotted peak in lower plots at ~2.7keV in **B**, **C**). Other peaks in the data and LUT have negligible intensity (**B**, **C**).

**Table S1:** Image sources.

| Fig. 1 | |
|---|---|
| 1A | HiRISE Color Basemap v5p1 NASA/JPL/University of Arizona |
| 1B | HiRISE Color Basemap v5p1 NASA/JPL/University of Arizona |
| 1C | NLF_0612_0721280802_722FDR_N0301172NCAM00709_0A1095J03 (*67*) |
| 1D | SI1_0612_0721292646_167FDR_N0301172SRLC00033_000095J02 (*69*) |
| 1E | NLF_0504_0711684404_926FDR_N0261222NCAM00709_0A0095J01 (*67*) |
| 1F | SIF_0504_0711696434_832FDR_N0261222SRLC02504_0000LMJ01 (*69*) |
| **Fig. 2** | |
| 2A | SC3_0614_0721480198_003FDR_N0301172SRLC10600_0000LMJ01 (*69*) |
| 2B | SC3_0513_0712520578_316FDR_N0261222SRLC16015_0000LMJ01 (*69*) |
| **Fig. 3** | |

| 3A | ZRF_0612_0721272628_348RAS_N0301172ZCAM03480_1100LMJ01 (*70*) |
| 3B | ZLF_0502_0711509150_010FDR_N0261222ZCAM07101_1100LMJ01 (*70*) This image is an oblique view of the outcrop. To reduce angular errors, and to align it with the other representations of the outcrop, it has been transformed with the following transformation matrix: [ 2.43726, 0.19372, 0.00017] [0.29979, 3.35840, 0.00033] [-1204.20588, -786.28054, 1.0] |
| **Fig. 4** | |
| 4A | SI1_0612_0721292646_167FDR_N0301172SRLC00033_000095J02 (*69*) |
| 4D | SC3_0614_0721480198_003FDR_N0301172SRLC10600_0000LMJ01 (*69*) |
| **Fig. 5** | SC3_0513_0712520578_316FDR_N0261222SRLC16015_0000LMJ01 (*69*) |
| **Fig. S3** | |
| S3A | NLF_0527_0713726823_925FDR_N0261222NCAM00712_0A0095J01 (*67*) |
| S3B | NLF_0626_0722520924_850FDR_N0301172NCAM00709_0A0095J01 (*67*) |
| S3C | CCFC0509_0712155363_000FDR_N0261222CACH00105_0A00LLJ01-s3 (*67*) |
| S3D | CCFC0516_0712775180_000FDR_N0261222CACH00105_0A00LLJ01-s3 (*67*) |
| S3E | CCF_0626_0722537762_729FDR_N0301172CACH00228_0M00LLJ01-s3 (*67*) |
| **Fig. S9** | SC3_0617_0721745135_414FDR_N0301172SRLC10600_0000LMJ01 (*69*) |
| **Fig. S10** | SC3_0513_0712520578_316FDR_N0261222SRLC16015_0000LMJ01 (*69*) |